\documentclass[prl,twocolumn,superscriptaddress,showpacs,floatfix]{revtex4}
\usepackage{graphicx}

\newcommand{\be}{\begin{equation}}
\newcommand{\ee}{  \end{equation}}
\newcommand{\ba}{\begin{eqnarray}}
\newcommand{\ea}{  \end{eqnarray}}
\newcommand{\ve}{\varepsilon}

\begin{document}

\title{Nuclear Excitation by a Zeptosecond Multi--MeV Laser Pulse}

\author{Hans A. Weidenm{\"u}ller}

\affiliation{Max-Planck-Institut f{\"u}r Kernphysik, D-69029 Heidelberg,
Germany}

\begin{abstract}

A zeptosecond multi--MeV laser pulse may either excite a ``plasma'' of
strongly interacting nucleons or a collective mode.  We derive the
conditions on laser energy and photon number such that either of these
scenarios is realized. We use the nuclear Giant Dipole Resonance as a
representative example, and a random--matrix description of the
fine--structure states and perturbation theory as tools.

\end{abstract}


\maketitle

{\it Purpose. Qualitative Considerations.} With the start of the
construction of ELI (the ``extreme light infrastructure'')~\cite{ELI}
or with extant ultra--intense laser facilites like NIF (if
reconfigured as femtosecond pulse systems)~\cite{NIF,Taj02} nuclear
spectroscopy using intense high--energy laser beams with short pulses
has become a realistic possibility. Indeed, it is envisaged to
generate in the decade ahead pulsed laser light with photon energies
of several MeV and pulse lengths of $10^{- 19}$ seconds by coherent
Thomson backscattering~\cite{Kie09,Wu10,Mou11}. This will be possible
provided that present intense experimental and theoretical efforts
will validate the concept of an electron mirror~\cite{Mou11}. These
very exciting developments call for a theoretical exploration of the
expected nuclear excitation processes. In the framework of the nuclear
shell model (a mean--field approach with a residual nucleon--nucleon
interaction), two scenarioss come to mind, distinguished by time
scales. (i) The time scale for the residual interaction is large
compared to the time scale for laser excitation of individual
nucleons. Then a single laser pulse containing $N \gg 1$ photons
excites many nucleons more or less simultaneously. The resulting
``plasma'' of excited and interacting nucleons (distantly similar to
the initial stage of a precompound reaction) is instable. Nucleons
excited above particle threshold with low angular momenta are emitted
instantaneously. The remainder of the system is equilibrated by the
residual interaction. Exciting questions are: What are the mean mass
number, the mean excitation energy and the mean angular momentum of
the resulting compound nucleus? How big are the spreads of these
quantities? Presumably it will be possible to study compound nuclei at
excitation energies and spin values not accessible so far. (ii) The
time scale for the residual interaction is sufficiently short compared
to the time between two successive photon absorption processes. Then
the nucleus relaxes after each photon absorption process. Single
photon absorption leads to a collective mode (typically the Giant
Dipole Mode), and multiple photon absorption within the same laser
pulse may lead to the formation of higher harmonics of that
mode. Thus, scenarios (i) and (ii) lead to extremely different forms
of nuclear excitation.

In this Letter we establish the time scales and the resulting
conditions on the mean photon energy $E_L$ and the number $N$ of
photons relevant for scenarios (i) and (ii). We do so by studying
scenario (ii) in detail. We show that scenarios (i) and (ii) both
occur for realistic choices of $E_L$ and $N$. We also show that
scenario (ii) is dominated by single--photon absorption.

We focus attention on dipole absorption, the dominant photon
absorption process in nuclei. The dipole mode $| 1 0 \rangle$ is the
normalized product of the dipole operator and the wave function $| 0
\rangle$ of the nuclear ground state. The dipole mode is not an
eigenstate of the nuclear Hamiltonian $H_{\rm nuc}$ and is spread over
the eigenstates $| \mu \rangle$ of $H_{\rm nuc}$ with eigenvalues
$E_\mu$, $\mu = 1, \ldots$. Gross features versus excitation energy
$E$ of that spreading are measured by the strength function $S(E) =
\sum_\mu \overline{ | \langle 1 0 | \mu \rangle |^2 \delta(E -
E_\mu)}$. The average (overbar) is taken over an energy interval large
compared to the average nuclear level spacing $d$. In the simplest
model adopted here, $S(E)$ has Lorentzian shape and is characterized
by two parameters~\cite{Har01}: The peak energy $E_{\rm dip} \approx
80 \ A^{-1/3}$ MeV (where $A$ is the nuclear mass), and the width
$\Gamma^\downarrow \approx 5$ MeV (the ``spreading width''). The
resulting broad peak of $S(E)$ is referred to as the Giant Dipole
Resonance (GDR). By the uncertainty relation, the time for the dipole
mode to spread over the eigenstates of $H_{\rm nuc}$ (the
``equilibration time'') is $\tau_{\rm eq} = \hbar /
\Gamma^\downarrow$.

The width $\Gamma_{\rm dip}$ for gamma decay of the GDR to the nuclear
ground state is estimated below and has a typical value of $5$ -- $10$
keV. With $N$ coherent photons in the laser pulse, the characteristic
time scale for photon absorption is $\tau_{\rm dip} = \hbar / (N
\Gamma_{\rm dip})$. One would expect that excitation of the GDR (as
opposed to scenario (i) considered above) dominates whenever $\tau_{\rm
dip} > \tau_{\rm eq}$, i.e., whenever $N \Gamma_{\rm dip} <
\Gamma^\downarrow$. That simple estimate is modified by two factors,
however. (i) For a short laser pulse with energy spread $\sigma$
(where we take $\sigma \approx 10$ keV corresponding to a pulse length
of $\approx 10^{-19}$ s), the Lorentzian shape of the GDR produces for
$E_L < E_{\rm dip}$ an additional factor $[\Gamma^\downarrow / (E_L -
E_{\rm dip})]^2$. (ii) The characteristic cubic dependence of
$\Gamma_{\rm dip}$ on energy yields an additional factor $(E_L /
E_{\rm dip})^3$. In total the criterion for collective excitation of
the GDR at energy $E_L$ reads $N < (E_L / E_{\rm dip})^3 [(E_L -
E_{\rm dip})^2 / (\Gamma_{\rm dip} \Gamma^\downarrow)]$. With
$\Gamma_{\rm dip} = 10$ keV, $\Gamma^\downarrow = 5$ MeV, $E_{\rm dip}
= 14$ MeV, $E_L = 7$ MeV that yields $N < 5 \times 10^3$. That bound
on $N$ is significantly larger than the bound $N < \Gamma^\downarrow /
\Gamma_{\rm dip} \approx 700$ obtained from the naive estimate and
shows that even for an intense laser pulse, excitation of the
collective GDR is a realistic alternative in nuclei to multiple
excitation of individual nucleons provided only that $E_L$ is
sufficiently far below $E_{\rm dip}$. Thus, varying both $N$ and $E_L$
provides the exciting opportunity to investigate scenarios (i) and
(ii) separately as well as the dynamical interplay between.

We support these qualitative arguments by calculating the
probabilities $P_1$ for single--quantum dipole excitation and $P_2$
for double--quantum dipole excitation as functions of $N, E_L$, and
$\sigma$. To calculate $P_2$ we use the Brink--Axel
hypothesis~\cite{Bri55,Axe62}. The hypothesis implies that single
excitation of the dipole mode may be followed either by double
excitation of that mode (i.e., formation of the second harmonic) or by
dipole excitation of the configurations mixed with the single dipole
mode. We account for both possibilities and show that for $\sigma \ll
\Gamma^\downarrow$ the contribution from the Brink--Axel mechanism
dominates and yields $P_2 = (1/2) P^2_1$. For values of $N$ and of
$E_L$ such that $P_1 \ll 1$ that relation implies that single photon
absorption is the dominant process even if $N \gg 1$. Our result
suggests that the probability for nuclear excitation by $n$--fold
dipole absorption may be approximately given by $P_n \approx 2^{- n}
P^n_1$.  That would imply that in the regime where our approximations
apply ($P_1 < 1/2$ or so) multiple collective nuclear excitation is
unlikely.

For the complex configurations that mix with the single or double
dipole modes, we use a random--matrix model. Every such model is based
upon the implicit assumption that the equilibration time (here
$\tau_{\rm eq}$) is short compared to the time scale of the physical
process of interest (here $\tau_{\rm dip}$). Our use of random--matrix
theory is justified if the above--mentioned conditions for collective
excitation of the GDR are met. We also use perturbation theory to
calculate $P_1$ and $P_2$.  That is justified if $P_1$ and $P_2$ are
sufficiently small compared to unity. The resulting constraint is the
same as for the use of random--matrix approach itself.

{\it Hamiltonian.} We write the total Hamiltonian as
\be
{\cal H}(t) = H_{\rm nuc} + H(t)
\label{1}
\ee
where $H(t)$ stands for the time--dependent interaction with the laser
pulse. In constructing $H_{\rm nuc}$ we are guided by the following
qualitative picture~\cite{Gu01}.  In a closed--shell nucleus, the
dipole mode $| 1 0 \rangle$ is a superposition of one--particle
one--hole (1p 1h) excitations. That mode is embedded in a sea of 2p 2h
excitations $| 0 k \rangle$ where $k = 1, \ldots, K$ and $K \gg
1$. (Here and in what follows the first label of the state vector
counts the number of absorbed dipole quanta and the second enumerates
the states). The mixing of both kinds of excitations causes the dipole
mode to be distributed over the eigenstates of $H_{\rm nuc}$. The
absorption of a second dipole quantum may either lead from the dipole
mode $| 1 0 \rangle$ to the double dipole mode $| 2 0 \rangle$ (a 2p
2h state), or it may lead from one of the 2p 2h states $| 0 k \rangle$
to the dipole mode $| 1 k' \rangle$ of that same state (a 3p 3h
state). The double dipole mode $| 2 0 \rangle$ is similarly embedded
in a sea of 3p 3h states $| 0 \alpha \rangle$ with $\alpha = 1,
\ldots, L$. All of the states $| 1 k' \rangle$ are embedded in a sea
of 4p 4h states $| 0 \rho \rangle$ where $\rho = 1, \ldots, M$ and $M
\gg K$. The residual interaction of the nuclear shell model mixes
these configurations, and both the double dipole mode and the states
$| 1 k' \rangle$ are spread out over the eigenstates of $H_{\rm
nuc}$. In modeling this qualitative picture we disregard the fact that
single or double dipole excitation may populate states with different
spin and isospin values. $H_{\rm nuc}$ is accordingly schematically
written in matrix form as follows.
\be
H_{\rm nuc} = \left( \matrix{ E_0 & 0 & 0 & 0 & 0 & 0 & 0 \cr 0 & E_1
& V_{1 l} & 0 & 0 & 0 & 0 \cr 0 & V_{k 1} & \tilde{H}^{(1)}_{k l} & 0
& 0 & 0 & 0 \cr 0 & 0 & 0 & E_2 & V_{2 \beta} & 0 & 0 \cr 0 & 0 & 0 &
V_{\alpha 2} & \tilde{H}^{(2)}_{\alpha \beta} & 0 & 0 \cr 0 & 0 & 0
& 0 & 0 & \tilde{{\cal H}}_{k' l'} & W_{k' \sigma} \cr 0 & 0 & 0 & 0
& 0 & W_{\rho l'} & \tilde{h}_{\rho \sigma} \cr} \right) \ .
\label{2}
\ee Here $E_0$ is the energy of the nuclear ground state, while $E_1$
and $E_2$ are the mean excitation energies of the single and of the
double dipole modes. For simplicity we use a harmonic--oscillator
picture so that $E_2 - E_1 = E_1 - E_0 = E_{\rm dip}$. Moreover we put
$E_0 = 0$.  The real matrix elements $V_{1 l}$ mix the dipole mode
with the 2p 2h states $| 0 l \rangle$. These are governed by the
$K$--dimensional Hamiltonian matrix $\tilde{H}^{(1)}_{k
l}$. Similarly, the matrix elements $V_{2 \beta}$ mix the double
dipole mode with the 3p 3h states $| 0 \beta \rangle$. These are
governed by the $L$--dimensional Hamiltonian matrix
$\tilde{H}^{(2)}_{\alpha \beta}$. We write $\tilde{H}^{(1)}_{k l} =
E_1 \delta_{k l} + H^{(1)}_{k l}$ and $\tilde{H}^{(2)}_{\alpha \beta}
= E_2 \delta_{\alpha \beta} + H^{(2)}_{\alpha \beta}$ and assume that
both $H^{(1)}_{k l}$ and $H^{(2)}_{\alpha \beta}$ are random matrices,
members of the Gaussian Orthogonal Ensemble (GOE), with no
correlations between the elements of $H^{(1)}_{k l}$ and of
$H^{(2)}_{\alpha \beta}$. The spectra of $E_1 \delta_{k l} +
H^{(1)}_{k l}$ and of $E_2 \delta_{\alpha \beta} + H^{(2)}_{\alpha
\beta}$ both have the shape of a semicircle centered at $E_1$ and
$E_2$, respectively. The last diagonal block in Eq.~(\ref{2})
describes similarly the mixing of the states $| 1 k' \rangle$ with the
4p 4h states $| 0 \rho \rangle$. We write $\tilde{{\cal H}}_{k' l'} =
E_2 \delta_{k' l'} + {\cal H}_{k' l'}$ and $\tilde{h}_{\alpha \beta} =
E_2 \delta_{\alpha \beta} + h_{\alpha \beta}$. We implement the
Brink--Axel hypothesis by putting ${\cal H} = H^{(1)}$. Again, the
$M$--dimensional matrix $h_{\rho \sigma}$ is assumed to be a member of
the GOE. We calculate the excitation probabilities $P_1$ and $P_2$ as
ensemble averages for $K, L, M \to \infty$. In that limit, the
spreading widths of the single and double dipole mode and of each of
the states $| 1 k' \rangle$ are given by the generic
expression~\cite{Wei09} $\Gamma^\downarrow = 2 \pi v^2 \rho$ where
$v^2$ stands for the mean square of the relevant mixing matrix
elements and $\rho$ for the mean level density in the center of the
semicircle. To avoid unneccessary complexity we assume that all
spreading widths have the same value $\Gamma^\downarrow$. That
schematic picture can be refined if the need arises. We disregard the
fact that the states excited by gamma absorption may decay by particle
or by gamma emission. That is justified because the time scales
associated with such decay are orders of magnitude larger than both
$\tau_{\rm eq}$ and $\tau_{\rm dip}$.

For the time--dependent interaction Hamiltonian $H(t)$ we use a
semiclassical description (justified for $N \gg 1$) and write
\be
H(t) = \sqrt{N} g(t) H_{\rm dip} \ .
\label{3}
\ee
Here $H_{\rm dip}$ is the time--independent electromagnetic
interaction operator for a single--photon dipole transition. The
factor $\sqrt{N}$ accounts for the presence of $N \gg 1$ photons and
the ensuing factor $N$ in the transition rate. The dimensionless
function $g(t)$ describes the time dependence of the short laser
pulse. We use the ansatz
\be
g(t) = \exp [ - \sigma^2 t^2 / (2 \hbar^2) - i \omega_L t ] \ .
\label{4}
\ee
Fourier transformation of $g(t)$ shows that the mean energy of the
laser pulse is $E_L = \hbar \omega_L$, the spread in energy has width
$\sigma$. Actually the interaction $H_{\rm dip}$ depends on energy,
too, via the wave number $k$. For $\sigma \approx 10$ keV we may put
$k \approx k_L$ where $k_L = E_L / (\hbar c)$.

In the scheme of Eq.~(\ref{2}) the non--zero matrix elements of the
dipole operator are $\langle 1 0 | H_{\rm dip} | 0 \rangle$, $\langle
2 0 | H_{\rm dip} | 1 0 \rangle$, and $\langle 1 k' | H_{\rm dip} | 0
k \rangle$. We use the Brink--Axel hypothesis to write $\langle 1 k' |
H_{\rm dip} | 0 k \rangle = \delta_{k k'} \langle 1 k | H_{\rm dip} |
0 k \rangle$. We assume that all non--zero matrix elements of the
dipole operator have the same value written as $\langle H_{\rm dip}
\rangle$. That corresponds to a harmonic--oscillator approximation. To
estimate $\langle H_{\rm dip} \rangle$, we write the Hamiltonian
$H_{\rm int}$ describing the interaction with the electromagnetic
field in Coulomb gauge as $H_{\rm int} = - (1 / c) \vec{j} \vec{A}$.
Here $\vec{j}$ is the current and $\vec{A}$ the vector potential. In
our time--dependent approach the latter has the form of a wave packet,
\be
\vec{A}(\vec{x}, \Omega, t) = \alpha \int {\rm d} \omega \exp[ - i
\omega t ] \ \tilde{g}(\omega) \exp [ i \vec{k} \vec{r} ] \ \vec{\chi}
\ .
\label{6}
\ee
The unit vector $\vec{\chi}$ describes the polarization, $\Omega$
indicates the direction of the vector $\vec{k}$, and $k =
\sqrt{\vec{k}^2}$ and $\omega$ are related by $k = \omega / c$. The
function $\tilde{g}$ is the Fourier transform of $g(t)$ in
Eq.~(\ref{4}). We determine the normalization constant $\alpha$ from
the reqirement that the energy carried by $\vec{A}$ be equal to
$E_L$. We use the dipole approximation. That yields $\alpha^2 =
(\sigma E_L) / (\pi^{1/2} \hbar c)$.  Quantization of the
electromagnetic field for individual quanta that have the form of the
wave packet~(\ref{6}) yields for the energy density the expression
$n(E) = 1 / (4 \pi^{3/2} \sigma)$. From Fermi's golden rule, the total
width for dipole decay is $\Gamma_{\rm dip} = 2 \pi n(E_L) | \langle
H_{\rm dip} \rangle |^2$. Thus,
\be
| \langle H_{\rm dip} \rangle | = \sqrt{2 \pi^{1/2} \Gamma_{\rm dip}
\sigma} \ .
\label{8}
\ee
For the dipole width we use the Weisskopf estimate, $\Gamma_{\rm dip}
= \frac{3}{4} \frac{e^2}{\hbar c} (k R)^2 E_L$. With $R = 3 \times
10^{-13}$ cm and $E_L = 15$ MeV that gives $\Gamma_{\rm dip} \approx
10$ keV, so that $| \langle H_{\rm dip} \rangle | \approx 10$ keV,
too. A somewhat larger value for $\Gamma_{\rm dip}$ results when the
Thomas--Reiche--Kuhn sum rule is taken into account. Here we are
interested in order--of--magnitude estimates only, however.

{\it Perturbation Expansion.} We solve the time--dependent
Schr{\"o}dinger equation in the interaction representation where the
perturbation has the form
\be
\tilde{H}(t) = \exp [ i H_{\rm nuc} t / \hbar ] \ H(t) \ \exp [ -
i H_{\rm nuc} t / \hbar ] \ .
\label{11}
\ee
We assume that at time $t = - \infty$ the nucleus is in the ground
state $| 0 \rangle$. We determine perturbatively the probabilities
$P_1$ and $P_2$ that at time $t = + \infty$ one or two dipole quanta
have been absorbed.

At $t = + \infty$, the probability amplitude for occupation of the
state $| 1 0 \rangle$ reached after single--dipole absorption is
\ba
&& b_1 = \frac{1}{i \hbar} \langle 1 0 | \int_{- \infty}^{+
\infty} {\rm d} t \ \tilde{H}(t) | 0 \rangle \nonumber \\
&& = \frac{\sqrt{N}}{i \hbar} \langle H_{\rm dip} \rangle 
\int_{- \infty}^{+ \infty} {\rm d} t \ g(t) \langle 1 0 | \exp [ i 
H_{\rm nuc} t / \hbar ] | 1 0 \rangle \ ,
\label{12}
\ea
and analogously (with $\langle 10 |$ in the last line replaced by
$\langle 0k |$) for $b_{0 k}$. The corresponding amplitudes for
occupation of the states $| 2 0 \rangle$, $| 0 \alpha
\rangle$ and $| 1 k' \rangle$, $| 0 \rho \rangle$ reached after
double--dipole absorption are denoted by $b_2$, $b_\alpha$, $b_{1 k'}$
and $b_{0 \rho}$. For example, we have
\ba
b_{0 \rho} &=& \bigg( \frac{1}{i \hbar} \bigg)^2 \langle 0 \rho |
\int_{- \infty}^{+ \infty} {\rm d} t_1 \ \tilde{H}(t_1 )\int_{-
\infty}^{t_1} {\rm d} t_2 \ \tilde{H}(t_2) | 0 \rangle \nonumber \\
&=& \bigg( \frac{\sqrt{N}}{i \hbar} \bigg)^2 \langle H_{\rm dip}
\rangle^2 \int_{- \infty}^{+ \infty} {\rm d} t_1 \ g(t_1) \int_{-
\infty}^{t_1} {\rm d} t_2 \ g(t_2) \nonumber \\
&& \times \sum_{l l'} \langle 0 \rho | \exp [ - i H_{\rm nuc}) t_1 /
\hbar ] | 1 l' \rangle \delta_{l l'} \nonumber \\
&& \qquad \times \langle 0 l | \exp [ i \{ H_{\rm nuc} (t_1 - t_2) \}
/ \hbar ] | 1 0 \rangle \ .
\label{13}
\ea
The average probabilities for single and double dipole absorption are,
thus, given by
\ba
P_1 &=& \bigg\langle | b_1 |^2 + \sum_k | b_{0 k}|^2 \bigg\rangle \ ,
\nonumber \\
P_2 &=& \bigg\langle | b_2 |^2 + \sum_\alpha | b_{0 \alpha}|^2
\nonumber \\
&& \qquad + \sum_{k'} | b_{1 k'} |^2 + \sum_\rho | b_{0 \rho} |^2
\bigg\rangle \ .
\label{14}
\ea
The big angular brackets indicate the ensemble average. The first
(last) two terms that contribute to $P_2$ are due to double excitation
of the dipole mode and to the Brink--Axel hypothesis, respectively.

{\it Averages.} By way of example we perform the ensemble average for
$P_1$ and focus attention on the sum of the squares of the
time-dependent matrix elements in Eqs.~(\ref{12}). Using completeness
and a simple identity we obtain for these
\ba
&& \bigg\langle \langle 1 0 | \exp [ i H_{\rm nuc} (t_1 - t_2) / \hbar
] | 1 0 \rangle \bigg\rangle = \nonumber \\
&& \int_{- \infty}^{+ \infty} {\rm d} \ve \  \exp [ i \ve (t_1 - t_2)
/ \hbar ] \bigg( \frac{1}{2 i \pi} \bigg\langle \langle 1 0 | \frac{1}
{\ve^- - H_{\rm nuc}} | 1 0 \rangle \nonumber \\
&& \qquad - \langle 1 0 | \frac{1}{\ve^+ - H_{\rm nuc}} | 1 0 \rangle
\bigg\rangle \bigg) \ .
\label{17}
\ea
We use Eq.~(\ref{2}) to write
\ba
&& \bigg\langle \langle 1 0 | \frac{1} {\ve^\pm - H_{\rm nuc}} | 1 0
\rangle \bigg\rangle \nonumber \\
&& = \bigg\langle \langle 1 0 | \frac{1} {\ve^\pm - E_{\rm dip} - V_1
(\ve^\pm - H^{(1)})^{- 1} V_1^\dag} | 1 0 \rangle \bigg\rangle
\nonumber \\
&& = \frac{1}{\ve - E_{\rm dip} \pm (i/2) \Gamma^\downarrow} \ .
\label{18}
\ea
Using Eq.~(\ref{4}) for $g(t)$ and carrying out the time integrals
(see Eqs.~(\ref{12})), we find that $\ve$ is confined to an interval
of size $\sigma$ around $E_L$. Since $\sigma \ll \Gamma^\downarrow$,
the argument of the expression in Eq.~(\ref{18}) can be taken at $\ve
= E_L$. The remaining integration can be done. With the help of
Eq.~(\ref{8}) that yields
\be
P_1 = \frac{2 \pi N \Gamma_{\rm dip} \Gamma^\downarrow}{(E_L - E_{\rm
dip})^2 + (1/4) (\Gamma^\downarrow)^2} \ .
\label{22}
\ee
The result~(\ref{22}) is intuitively appealing and clearly displays
the suppression factors ${\Gamma^\downarrow}^2 / (E_L - E_{\rm
dip})^2$ and $(E_L / E_{\rm dip})^3$ mentioned above that come into
play for $E_L < E_{\rm dip}$.

The calculation of $P_2$ proceeds similarly but is more involved. We
use operator identities such as
\ba
&& \langle 1 0 | \frac{1}{\ve^-_2 - H_{\rm nuc}} | 0 k \rangle 
\nonumber \\
&& = \langle 1 0 | \frac{1}{\ve^-_2 - E_{\rm dip} - V_1 (\ve^-_2 -
E_{\rm dip} - H^{(1)})^{-1} V_1^\dag} | 1 0 \rangle \nonumber \\
&& \qquad \qquad \qquad \times \langle 1 0 | V_1 \frac{1}{\ve^-_2 -
E_{\rm dip} - H^{(1)}} | 0 k \rangle \ .
\label{28}
\ea
That leads to products of terms each containing $H^{(1)}$ in the
denominator. We neglect the correlations between eigenvalues of
$H^{(1)}$ in different factors because such correlations extend over
an energy range measured in units of the mean level spacing $d$ while
the range of the terms in Eq.~(\ref{28}) is given by
$\Gamma^\downarrow \gg d$. For the last two terms in the second of
Eqs.~(\ref{14}) we obtain
\ba
&& \bigg\langle \sum_{k'} | b_{1 k'} |^2 + \sum_\rho | b_{0 \rho} |^2
\bigg\rangle = \frac{1}{2} P^2_1
\label{32}
\ea
with $P_1$ given by Eq.~(\ref{22}). The calculation of the first two
terms yields a contribution that in comparison to Eq.~(\ref{32}) is
small of order $\sigma / \Gamma^\downarrow$. Thus for all values of
$E_L$ the contribution to $P_2$ from double excitation of the dipole
mode is negligibly small in comparison with that from the Brink--Axel
mechanism in Eq.~(\ref{32}). As a result we find
\be
P_2 = \frac{1}{2} P^2_1 \ .
\label{25}
\ee
The factor $1/2$ in Eq.~(\ref{25}) is due to the time ordering in
Eqs.~(\ref{13}). Thus, we expect that for arbitrary positive integer
$n$ we have $P_n = 2^{- n} P^n_1$.

{\it Conclusions.} We have established the time scales and the
resulting values for mean photon energy $E_L$ and mean photon number
$N$ required for the realization of either of the two scenarios
mentioned in the Introduction. This was done with the help of a
random--matrix model for scenario (ii) for which we have calculated
the probabilities $P_1$ and $P_2$ for single and double nuclear dipole
absorption. Our assumptions and approximations require both $P_1$ and
$P_2$ to be small compared to unity. In that case, scenario (ii)
applies. Eq.~(\ref{22}) shows that in the tails of the GDR that
condition is easily met even for an intense laser pulse.  Ways of
detecting such collective nuclear excitation experimentally are
discussed in Ref.~\cite{Die10}. Double photon absorption is dominantly
due to the Brink--Axel mechanism (as opposed to double excitation of
the dipole mode). For $P_1$ small compared to unity, single--photon
absorption is the dominant mechanism while $P_2 \ll P_1$.

With increasing $N$, the time for dipole absorption $\tau_{\rm dip}$
eventually becomes small compared to the nuclear equilibration time
$\tau_{\rm eq}$, and the competition between collective excitation and
the formation of a strongly interacting nucleon ``plasma'' is decided
in favor of the latter. Eq.~(\ref{22}) shows that in the center of the
GDR ($E_L \approx E_{\rm dip}$), that will happen already for fairly
small values of $N \approx 10$ or so. As $N$ is increased, the process
spreads to the tails of the GDR. It is a challenge to attain a
theoretical understanding of scenario (i), and of the interplay
between both scenarios.

{\it Acknowledgments.} I thank P. Thirolf for valuable advice, D. Habs
for stimulating discussions, and A. Richter and B. Dietz for helpful
suggestions.

\end{document}